\documentclass[prc,twocolumn,showpacs]{revtex4}
\usepackage{graphicx}
\usepackage{dcolumn}
\usepackage{bm}

\begin{document}

\title{Signatures of shape transition in odd-A neutron-rich Rubidium isotopes}

\author{R. Rodriguez-Guzman$^{1}$, P. Sarriguren$^{1}$}
\author{L.M. Robledo$^{2}$}
\affiliation{
$^{1}$ Instituto de Estructura de la Materia, CSIC, Serrano
123, E-28006 Madrid, Spain \\
$^{2}$ Departamento  de F\'{\i}sica Te\'orica, M\'odulo 15,
Universidad Aut\'onoma de Madrid, 28049-Madrid, Spain
}

\date{\today}

\begin{abstract}
The isotopic evolution of the ground-state nuclear shapes and the
systematics of one-quasiproton configurations are studied in odd-$A$
Rubidium isotopes. We use a selfconsistent Hartree-Fock-Bogoliubov
formalism based on the Gogny energy density functional with two
parametrizations, D1S and D1M, and implemented with the equal
filling approximation. We find clear signatures of a sharp shape
transition at $N=60$ in both charge radii and spin-parity of the
ground states, which are robust, consistent to each other, and in
agreement with experiment. We point out that the combined analysis
of these two observables could be used to predict unambiguously
new regions where shape transitions might develop.
\end{abstract}


\maketitle

The study of the nuclear shape evolution as the number of nucleons 
changes is nowadays a highly topical issue in nuclear physics from
both theoretical and experimental points of view (see for example 
Refs. \cite{wood,bender,rodriguez,ours} and references therein).
Specially interesting are those situations where the nuclear
structure suffers drastic changes between neighbor nuclides. These
structural variations lead often to sudden changes of particular
nuclear properties that can be used as signatures of phase/shape 
transitions \cite{ours_plb,ours_last}.
This is the case of the neutron-rich isotopes with masses $A\sim 100$.
Intense experimental  \cite{urban,campbell,charlwood} and theoretical 
\cite{skalski,xu02,ours_plb,ours_last} efforts are being done to
better characterize the structural evolution of the ground and
excited states in this region of the nuclear chart. In particular,
we have recently studied in Refs. \cite{ours_plb,ours_last} the
structural evolution in even and odd  neutron-rich 
Sr, Zr, and Mo isotopes. In this work we concentrate on the study of 
neutron-rich odd-mass Rb isotopes that have received considerable
attention recently \cite{lhersonneau01,bucurescu07,hwang09,simpson10}
and that exemplify a general pattern where structural fluctuations
lead to observable effects.

Being an odd-$Z$ nucleus, the spin and parity of odd-$A$ Rb isotopes
are determined by the state occupied by the unpaired proton. The
spectroscopic properties of the various isotopes will be determined
by the one-quasiproton configurations that, in principle, are expected
to be rather stable against variations in the number of even neutrons.
However, as it is known in neighboring nuclei, approaching $N\sim 60$,
the isotopes become well deformed \cite{wood,ours_plb} and the abrupt
change in deformation induces signatures in nuclear bulk properties
like the two-neutron separation energies and the nuclear charge radii,
as well as in spectroscopic properties. In particular, the spin and
parity of the nuclear ground state might change suddenly from one
isotope to another, reflecting the structural change.

\begin{figure}[h]
\centering
\includegraphics[width=70mm]{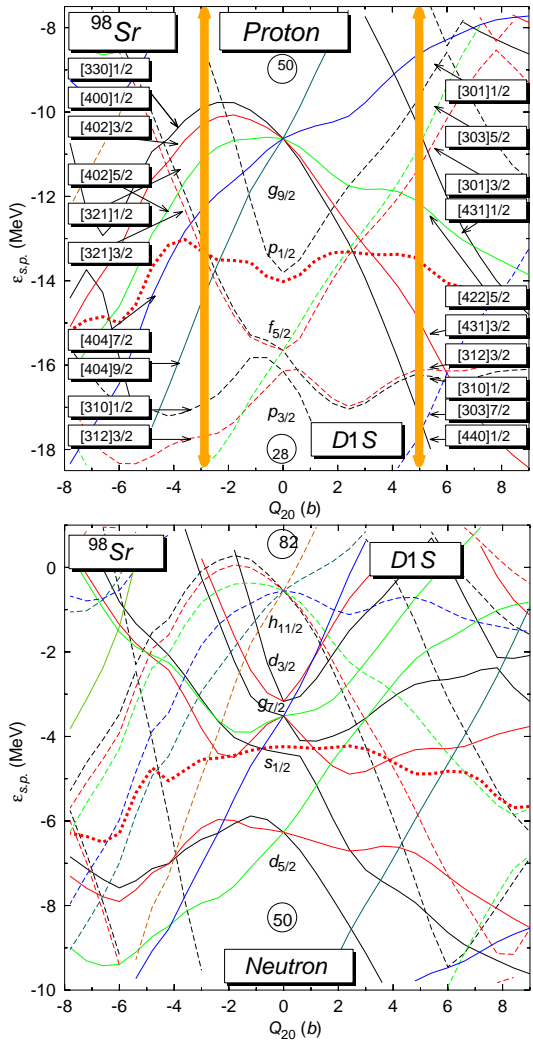}
\caption{(Color online) Single particle energies for protons and 
neutrons in $^ {98}$Sr as a function of the axial quadrupole moment 
$Q_{20}$. The Fermi level is depicted as a thick dashed red line. 
The results have been obtained with the Gogny-D1S EDF.}
\label{fig_spe}
\end{figure}

We analyze in this work the bulk and spectroscopic properties of
neutron-rich Rb odd isotopes within the selfconsistent
Hartree-Fock-Bogoliubov (HFB) approximation based on the finite
range and density dependent Gogny energy density functional (EDF)
\cite{gogny}. Previous studies in this region including triaxiality
\cite{ours_plb} suggest that the $\gamma$-degree of freedom would
not play a significant role in these isotopes and therefore, axial
symmetry is assumed as a selfconsistent symmetry in this study.

In addition to the well known D1S parametrization \cite{d1s} of
the Gogny-EDF, we also consider the most recent parametrization
D1M \cite{d1m}. From this comparison we evaluate not only the
robustness of our results, but we also explore the capability of
D1M to account for the phenomenology of odd-$A$ nuclei, not so
well studied yet. The description of the odd-$A$ nuclei involves
additional difficulties because the blocking procedure requires the
breaking of time-reversal invariance, making the calculations 
arduous \cite{duguet,bonneau}. In the present study we use the
equal filling approximation (EFA), a prescription widely used in
mean-field calculations to preserve the advantages of time-reversal
invariance. In this approximation the odd nucleon sits half into
a given orbital and half into its time-reversed partner. The
microscopic justification of the EFA is based on standard ideas of
quantum statistical mechanics \cite{perez}. The predictions arising
from various treatments of the blocking have been studied in 
Ref. \cite{schunck}, concluding that the EFA is sufficiently
precise for most practical applications. More details of our
procedure can be found in Ref. \cite{ours_last}.
 
\begin{figure}[h]
\centering
\includegraphics[width=70mm]{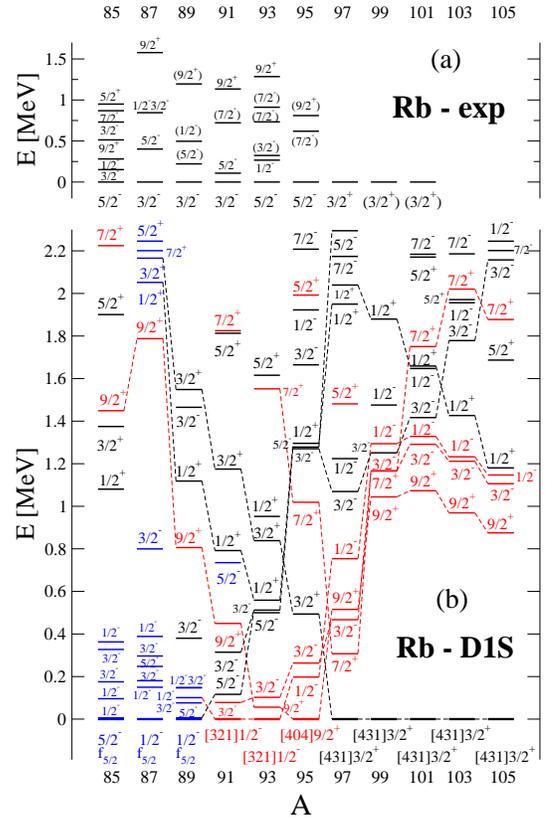}
\caption{(Color online) Experimental (a) excitation energies and spin-parity 
assignments \cite{lhersonneau01,bucurescu07,hwang09,simpson10,exp_ensdf}
compared with Gogny-D1S HFB-EFA results (b) for one-quasiproton states in
odd-$A$ Rb isotopes. Prolate configurations are shown by black lines, oblate
ones by red lines, and spherical ones by blue lines.}
\label{fig_rb}
\end{figure}

The proton and neutron single-particle energies (SPEs) are shown in 
Fig. \ref{fig_spe} as a function of the axial quadrupole moment $Q_{20}$
for the even-even $^{98}$Sr $(Z=38, N=60)$ isotope. Fermi levels are
plotted with thick dashed (red) lines. Asymptotic (Nilsson) quantum
numbers $[N,n_z,\Lambda]K^\pi$ are also shown in the proton case for the
$Q_{20}$ values where the energy minima are located (vertical arrows)
in both oblate and prolate sectors. As one can see, the valence protons
occupy the $N=3$ shell and, if deformed, they start to fill the $g_{9/2}$
orbitals coming down from the $N=4$ shell. Neutrons occupy states
belonging to the $N=4$ shell and approaching $N=60$ they start to
populate the $h_{11/2}$ intruder orbitals coming down from the $N=5$
oscillator shell at large deformations ($Q_{20}\sim 4-5$ b). These neutrons
polarize the protons and drive them to populate the $1g_{9/2}$ shell at
similar deformations. Thus, the underlying nuclear structure in this mass
region is very sensitive to the occupancy of these single-particle orbitals
and the result is a rapid change in the nuclear deformation and in the
spectroscopic properties as a function of both neutron and proton numbers. 

In particular, Rb isotopes approaching $N=60$ will jump abruptly from
slightly deformed oblate or prolate shapes to strongly prolate shapes.
As a consequence, the unpaired proton that finally determines the spin
and parity of the whole nucleus in Rb isotopes, will jump from the 
$[303]5/2^-$ at low deformation to the $[431]3/2^+$ orbital at large 
deformation. This is fully consistent with
the Jahn-Teller picture, which states that nuclei avoid regions with
high single-particle level densities around the Fermi level, thus
favoring deformation where the energy gaps are higher. One sees large
energy gaps at large prolate deformations for the  $[431]3/2^+$ orbital.
This is also consistent with the Federman-Pittel mechanism \cite{fp},
which points to the $T=0$ neutron-proton interaction as the driven force
toward quadrupole deformation. This force is particularly intense between
spin-orbit partners (neutron $1g_{7/2}$ and proton $1g_{9/2}$ in our case),
as well as between orbitals with the same radial quantum numbers and
large orbital angular momenta differing by one unit $n_p=n_n$ and 
$\ell_p=\ell_n\pm 1$ (neutron $1h_{11/2}$ and proton $1g_{9/2}$ in our case).

\begin{figure}[h]
\centering
\includegraphics[width=70mm]{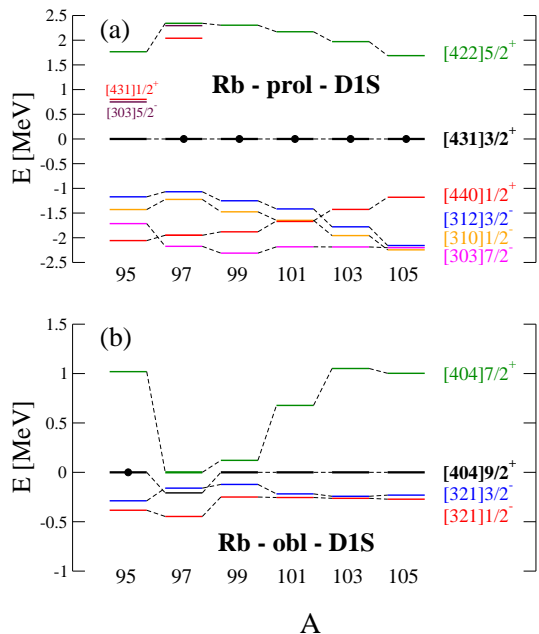}
\caption{(Color online) Gogny-D1S excitation energies of single-quasiproton
prolate (a) and oblate (b) states in Rb isotopes. Hole states are plotted
below zero energy and particle states are plotted above. The absolute
ground states are indicated with a circle.}
\label{fig_d1s}
\end{figure}

\begin{figure}[h]
\centering
\includegraphics[width=70mm]{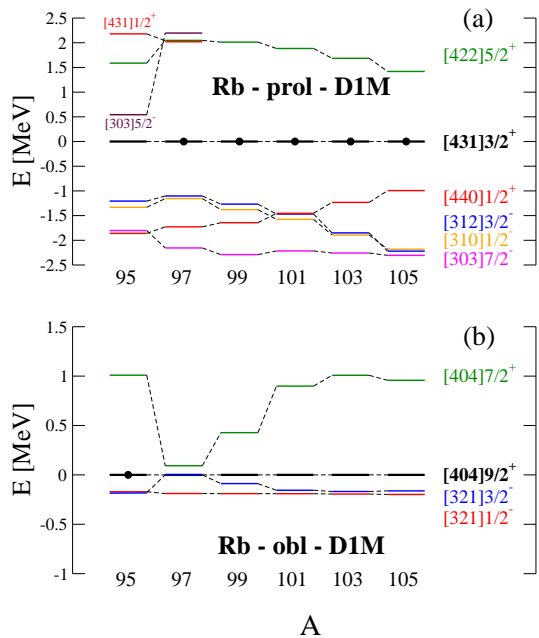}
\caption{(Color online) Same as in Fig. \ref{fig_d1s}, but for Gogny D1M.}
\label{fig_d1m}
\end{figure}

\begin{figure}[h]
\centering
\includegraphics[width=70mm]{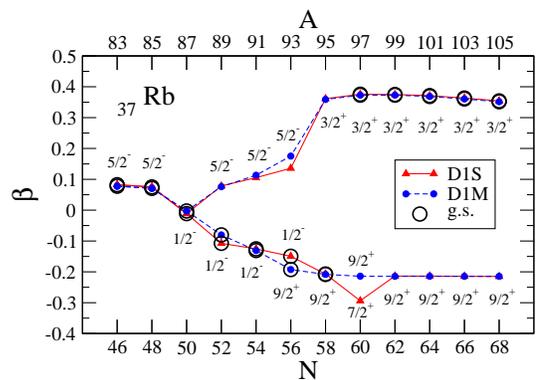}
\caption{(Color online) Isotopic evolution of the quadrupole deformation
parameter $\beta$ of the energy minima obtained from Gogny D1S and D1M
calculations. }
\label{fig_def}
\end{figure}

In Fig. \ref{fig_rb} we can see the experimental excitation energies and
spin-parity assignments (a) in odd-$A$ neutron-rich Rb isotopes.
They are compared to the one-quasiproton states predicted by the Gogny-D1S
HFB-EFA calculation (b). The excited states in a given isotope are referred
to the corresponding ground state, regardless its shape. Prolate configurations
in our calculations are shown by black lines, oblate ones by red lines, 
and spherical ones by blue lines. The quasiparticle states are labeled by
their $K^\pi$ quantum numbers. The most important configurations are
joined by dashed lines following the isotopic evolution. In addition,
the ground states are labeled by their asymptotic quantum numbers.

Experimentally \cite{lhersonneau01,bucurescu07,hwang09,simpson10,exp_ensdf}
one observes $J^\pi=3/2^-$ ground states in $^{87,89,91}$Rb, then $5/2^-$ states
in $^{93,95}$Rb, and finally  $3/2^+$ states in the heavier isotopes. The $9/2^+$
states appear all the way up to $^{95}$Rb as excited states. The most striking
feature observed is a jump at $N=60\, (A=97)$ from $5/2^-$ to $3/2^+$ ground
states.

The theoretical interpretation of these features can be understood from the
analysis of our results in the lower panel (b). Rb isotopes evolve from
spherical shapes in $^{85-89}$Rb around $N=50$ with the spherical $f_{5/2}$, 
$p_{3/2}$, and $p_{1/2}$ shells involved, to slightly deformed shapes (oblate
and prolate almost degenerate) in $^{91-95}$Rb, and finally to well deformed
prolate shapes in $^{97-105}$Rb. In the lighter isotopes the three spherical
shells mentioned above are very close in energy and thus, we can see all
the split $K^\pi$ levels coming from them at very close low-lying excitation
energies. In the case of $^{91-95}$Rb isotopes we obtain oblate ground states
with both oblate and prolate low-lying excited states very close in energy.
For instance, we see that the experimental $3/2^-$ ground states in 
$^{89,91}$Rb appear in the calculation as excited states at less than 0.1 MeV. 
In the case of $^{93,95}$Rb the experimental ground states $5/2^-$ appear 
as prolate excited states, while the calculations produce oblate ground
states. The ground states of the heavier isotopes are predicted to be
$3/2^+$ in agreement with experiment. The calculations also predict 
excited low-lying $9/2^+$ states with an oblate nature, which are observed
experimentally.

In the next figures we compare the spectroscopic properties of the Gogny
D1S and D1M parametrizations. Contrary to Fig. \ref{fig_rb}, in Fig. 
\ref{fig_d1s} for D1S and in Fig. \ref{fig_d1m} for D1M, we have separated
the prolate (a) and oblate (b) states and have plotted the most relevant
hole states below zero energy and the particle states above. The absolute
ground states, either oblate or prolate, are indicated with a circle.
Specifically, in the figures corresponding to the prolate shape, we have
plotted the evolution of the $3/2^+(g_{9/2})$ states, which are ground states
for $^{97-105}$Rb. In the upper region we find the $5/2^+(g_{9/2})$, while in
the lower region we have the $1/2^+(g_{9/2})$, $3/2^-(p_{3/2})$, $1/2^-(f_{5/2})$, 
and $7/2^-$ coming up from the $f_{7/2}$ shell. In the oblate case (b) we
can see the $9/2^+(g_{9/2})$ state, which is ground state in $^{95}$Rb, the 
$7/2^+(g_{9/2})$ as particle state and the $1/2^-$ and $3/2^-$ from $f_{5/2}$
as hole states.

Very similar results are obtained from both parametrizations, which answer
our original questions about robustness of the calculations and reliability
of D1M. The only difference worth mentioning is that D1M produces slightly
lower excited states. This feature can be understood from its larger
effective mass that makes the single particle spectrum somewhat more dense
with D1M.

To further illustrate the role of deformation and spin-parity assignments
in the isotopic evolution, we display in Fig. \ref{fig_def} the axial
quadrupole deformation $\beta$ of the energy minima as a function of $N$
with both D1S and
D1M parametrizations. The deformation of the ground state for each isotope
is encircled. We can see that starting at the semi-magic isotope with
$N=50$ we get two minima in the prolate and oblate sectors. These are
$5/2^-$ in the prolate case and $1/2^-$ (or $9/2^+$ very close in energy)
in the oblate case. The calculations with the two forces predict oblate
ground states, but with prolate solutions which are very close in energy
and that could perfectly become ground states for slightly different type
of calculations.
Thus, in $^{89}$Rb the $1/2^-$ and $5/2^-$ are practically degenerate
(see Fig. \ref{fig_rb}). In  $^{91}$Rb ($^{93}$Rb) they are separated by
less than 0.2 (0.5) MeV. The experimental information seems to favor
prolate solutions since $5/2^-$ ground states are observed. The ground
states of heavier isotopes starting at $^{97}$Rb are $3/2^+$ in agreement
with experiment and strongly prolate. The $9/2^+$ oblate configurations
appear all the way toward heavier isotopes.

\begin{figure}[h]
\centering
\includegraphics[width=70mm]{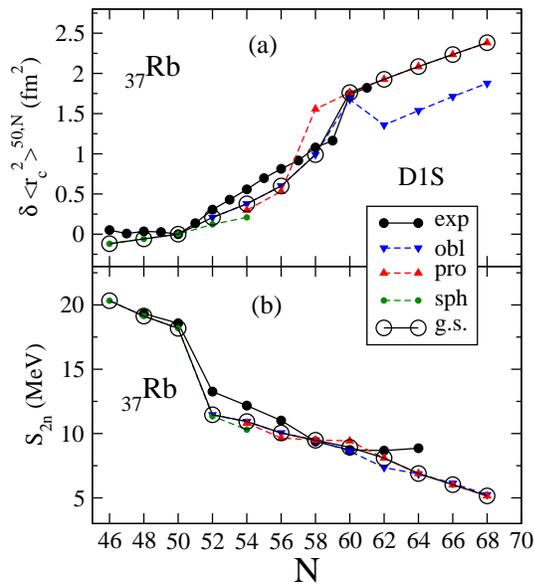}
\caption{Gogny-D1S HFB results for $\delta \langle r^2_c \rangle$ (a) and 
$S_{2n}$ (b) in odd-$A$ Rb isotopes compared to experimental data from Ref. 
\cite{audi,rahaman07} for masses and from Ref. \cite{thibault}
for radii. Results for prolate, oblate, and spherical minima are displayed with
different symbols (see legend). Open circles correspond to ground-state
results.}
\label{fig_r_d1s}
\end{figure}

\begin{figure}[h]
\centering
\includegraphics[width=70mm]{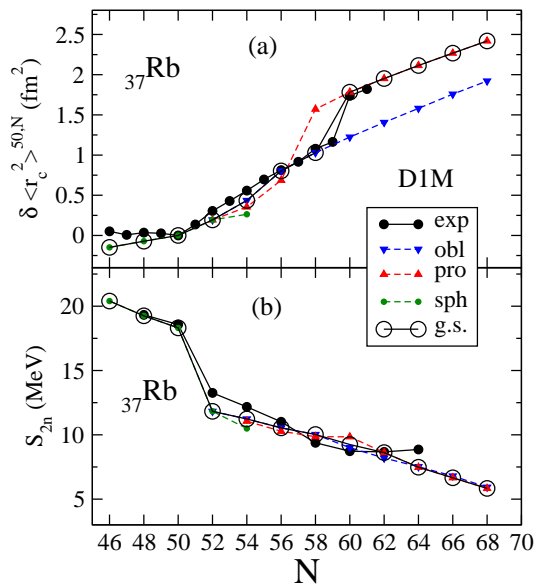}
\caption{Same as in Fig. \ref{fig_r_d1s}, but for Gogny D1M.}
\label{fig_r_d1m}
\end{figure}

Finally, we show in Fig. \ref{fig_r_d1s} for D1S and in Fig. \ref{fig_r_d1m} 
for D1M, the charge radii differences (a) defined as $\delta \langle r^2_c 
\rangle ^{50,N}= \langle r^2_c \rangle ^N - \langle r^2_c \rangle ^{50}$, 
calculated with the same corrections as in Ref. \cite{ours_plb}, and the
two-neutron separation energies $S_{2n}$ (b). We compare our results for
$\delta \langle r^2_c \rangle $ with isotope shifts from laser spectroscopy
experiments \cite{thibault} and our results for $S_{2n}$ with
masses from Refs. \cite{audi,rahaman07}. 
The results for $S_{2n}$ agree in general with the measurements and are
very similar between D1S and D1M. A somewhat better agreement is found
with D1M at the beginning of the shell ($N>50$). In any case,
it would be interesting to see whether the incipient experimental tendency
to raise the $S_{2n}$ values in the heavier isotopes and not predicted by
the calculations, still persists in more exotic nuclei. In this respect,
we should remark that for the heavier Rb isotopes the mass data are not
completely reliable because they are based on $\beta$-endpoint measurements.
It has been shown that this type of indirect measurements systematically
tend to underestimate the $Q$-values leading to very strong binding.
This problem has been discussed in Ref. \cite{hager} for neutron-rich Y and
Nb isotopes. Thus,
it would be very helpful to extend and improve mass measurements reducing
the still large uncertainties in neutron-rich Rb isotopes.

The nuclear charge radii plotted in the upper figures show a jump at $N=60$,
where the radius suddenly increases. This experimental observation is well
reproduced in our calculations, where the encircled ground states show that
the jump occurs between $N=58$ and $N=60$. This jump correlates well with
the one observed in the spin-parity of the ground states displayed in
Fig. \ref{fig_rb} from  $5/2^-$ to $3/2^+$. Theoretically, the jump in 
$\delta \langle r^2_c \rangle $ could be understood as the result of the
change from an oblate to a prolate shape. But it could also be interpreted
as a jump from a slightly prolate shape to a well deformed prolate nucleus,
because the oblate and prolate shapes in the transitional nuclei $N=54-58$
have very similar absolute values of deformation (see Fig. \ref{fig_def}).
This similarity gives raise to practically the same charge radii, as it
can be seen by comparing the prolate (upward red triangles) and oblate
(downward blue triangles) radii in this region.
The only difference worth mentioning between D1S and D1M is the oblate
radii at $N=60$ which corresponds to different configurations ($7/2^+$
in D1S and $9/2^+$ in D1M) and different deformations
(see Fig. \ref{fig_def}).

Then, we see that contrary to $S_{2n}$, which does not signal any clear 
signature in Rb isotopes, $\delta \langle r^2_c \rangle $ clearly indicates
the existence of a shape transition in agreement with experiment and well
correlated to the spin-parity jump. It would be also very interesting to
extend the isotope shifts measurements, which are rather old, to more
exotic isotopes to confirm also the prolate stabilization.

In summary, we have studied the shape evolution in odd-$A$ Rb isotopes
from microscopic selfconsistent Gogny-EDF HFB-EFA calculations.
We have analyzed various sensitive nuclear observables, such as 
two-neutron separation energies, charge radii, and the spin-parity
of ground states in a search for signatures of shape transitions.
We have found that, although the masses are not very sensitive to
these shape changes in Rb isotopes, the charge radii and the
spin-parity are. In addition,
the signatures found are all consistent to each other and point
unambiguously to the existence of a shape transition. The correlations
between all of these signatures are useful to establish a protocol to
look for shape transitions in the future in other regions of the
nuclear chart and to predict and identify them. We have also analyzed
the isotopic evolution of the one-quasiproton configurations and have
compared the predictions of two different Gogny-EDF parametrizations,
demonstrating the robustness of our calculations.

The experimental information available in neutron-rich Rb isotopes and
in general in this mass region is still very limited. It would be highly
desirable to extend the experimental programs for mass, charge radii, and
spectroscopic measurements to these exotic regions at existing facilities
like ISOLTRAP \cite{mukherjee} at ISOLDE/CERN and IGISOL \cite{jokinen}
at the University of Jyv\"askyl\"a, or at future ones like MATS and
LaSpec \cite{rodriguez} at the FAIR facility, where we can learn much
about structural evolution in nuclear systems.

The theoretical efforts should be also pushed forward by improving the
formalism including triaxial degrees of freedom in those regions where
this can be an issue or dealing with the odd systems with exact blocking
treatments.
The quality of our mean field description could be improved from
configuration mixing calculations in the spirit of the Generator
Coordinate Method with the quadrupole moment as generator coordinate.
Nevertheless, such an approach for odd nuclei is out of the scope
of the present study. It is already known \cite{bender_08} that such
a configuration mixing reduces the jump in $S_{2n}$ predicted in
pure mean field approximations when crossing shell closures,
improving the agreement with experiment. This could be particularly
relevant for the light isotopes considered in the present study,
where the spherical minima are rather shallow. For heavier isotopes,
the two minima, oblate and prolate, are separated by spherical barriers
of about 3 MeV and appear about 1 MeV apart. The effect here is not
expected to be significant because a single shape would be enough
to account for the properties studied.
A simple two-state mixing model could be used to mix the intrinsic
deformed configurations into the physical states \cite{wood}.
Assuming that the wave functions are localized at largely
different values of the deformation coordinate, one can neglect
the cross terms. The final effect would be to place the value of
the radius in between the oblate and prolate values according
to the mixing amplitude of those intrinsic configurations into
the physical wave function. The consequence would be a slightly
smoother transition.

\noindent {\bf Acknowledgments}

This work was supported by MICINN (Spain) under research grants 
FIS2008--01301, FPA2009-08958, and FIS2009-07277, as well as by 
Consolider-Ingenio 2010 Programs CPAN CSD2007-00042 and MULTIDARK 
CSD2009-00064.
We thank Prof. J. \"Aysto and Prof. P.M. Walker for valuable
suggestions and discussions.

\end{document}